# Lost in k-Space: An Open-Source MR-Physics Escape Room

Sabine Melanie Räuber[1,2], Marta Brigid Maggioni[1,2], Francesco Santini[1,2]

[1] Basel Muscle MRI (BAMM), Department of Biomedical Engineering, University of Basel, Basel, Switzerland.

[2]Radiology, Division of Radiological Physics, University Hospital of Basel, Basel, Switzerland.

sabine.raeuber@unibas.ch, martabrigid.maggioni@unibas.ch , francesco.santini@unibas.ch

**Corresponding author**: Francesco Santini,

*E-mail: francesco.santini@unibas.ch*

**Keywords:** Magnetic Resonance Imaging, Physics, Professional Education, Gamification

# Abstract

**Introduction.** Operating an MR scanner for technical and clinical research requires multidisciplinary competencies beyond MR physics, including safety management and teamwork. Gamification, particularly educational escape rooms, have been associated with improved motivation, engagement and knowledge retention in health professional education.

**Materials and Methods.** We developed an MR-physics-themed educational escape room for the 41st Annual Meeting of the ESMRMB. Designed for teams of four with a time limit of 25 minutes, the room reproduced the atmosphere of an MR control room. The five puzzles covered the Larmor equation, sequence composition, MR safety and acoustical identification of MR sequences. Custom ESP32 electronics allowed the puzzles to communicate in real time with each other and with the game master

**Results.** The room ran without technical interruptions and was played by 39 teams (approximately 160 participants); 12 solved it (escape rate 31%), with a median solution time of 21 minutes and 53 seconds. Early-career researchers acted as game masters.

**Discussion.** The escape rate aligns with comparable activities targeting scientific audiences. Difficulty can be tuned by adjusting puzzle obscurity, component availability, required prior knowledge and mental leaps. A modified version could usefully supplement mandatory MR safety training. Code, schematics, machining files and documentation are released as open source.

# Introduction

MR Physics concepts tend to play a central role in the daily occupational life of our readers, and of the members of the European Society for Magnetic Resonance in Medicine and Biology (ESMRMB). Operating an MRI scanner for technical and clinical research requires a broad range of multidisciplinary skills. While these skills are rooted in the physical principles of nuclear magnetic resonance (NMR), mastering them extends far beyond physics to include scanner operation, pulse sequence optimization, patient handling, image quality assessment, safety management, and interdisciplinary team communication. Learning and teaching these core competencies is integral to our work.

To develop these theoretical and practical competencies, traditional teaching methods include lectures, practical demonstrations, and hands-on scanner sessions. Gamification [1] has emerged as a complementary educational approach that incorporates game-based elements into learning activities to promote active participation and engagement. One popular gamification strategy is educational escape rooms [2]. Escape rooms are live-action, team-based games in which players must overcome multiple challenges within a limited timeframe in order to unlock the room door and successfully escape. The implementation of escape room methodologies has been observed across various academic disciplines, including STEM (Science, Technology, Engineering and Mathematics) [3] and health professional education [4, 5].

In the medical field, escape rooms have been associated with improved learner motivation, increased engagement, and enhanced retention of theoretical knowledge as well as promoting teamwork [6–11].

Translating this concept to magnetic resonance, we developed an MRI-themed educational escape room, based on practical puzzles and custom-built electronics, for the 41st annual conference of the European Society for Magnetic Resonance in Medicine and Biology (ESMRMB) in Marseille (France). The escape room was available to be played throughout all three days of the main congress (Thursday to Saturday) for all registered participants. In this paper, we present the game's design and release all code and schematics open-source, offering a ready-to-use blueprint for future medical physics outreach and teaching.

# Materials and Methods

## Design concepts

The escape room was designed to be solved in 25 minutes by teams of 4 people, although smaller or larger teams were also accepted to play. The scenario of the room was to simulate a situation in which the player needed to safely execute an MRI scan, and the layout of the room mimicked an MRI scanner control room. The concepts applied in the room were the following:

- basics of MR physics (Larmor frequency)
- basics of MR sequence design (schematic representation of radiofrequency pulses and trapezoidal gradients)
- basics of MR safety
- practical scanning experience (recognizing sequences from their noise at the scanner)

The room consisted eventually of five interoperating puzzles, which could be solved in a partially non-sequential fashion as depicted in Figure 1.

## Puzzles and solutions

The main puzzles and accompanying solutions are as follows:

## Puzzle 1: Amplifier cabinet

Positioned in the room is a cabinet containing two drawers. The unlocked drawer is marked "Gradient amplifier," whereas the locked drawer is marked "RF amplifier - 1H@3T." Players deduce the 4-digit lock code by calculating the approximate Larmor frequency of hydrogen at 3T using a given table of (approximate) gyromagnetic ratios. Once opened, the drawer contains wooden tiles, on which basic sequence building blocks symbols are drawn (RF and gradient pulses). Also in the same drawer, an item labeled "ferromagnetic detector" is found.

## Puzzle 2: Sequence composer

The "sequence composer" puzzle consists of a wooden support, on top of which the tiles found in puzzle one need to be arranged to form the schematic for a spin-echo MR sequence. This particular sequence is hinted at by the presence of two "RF pulse" tiles of different amplitudes and of a tile with crusher gradients (Figure 1). Three LEDs are also present on the wooden structure: a red one labeled "MR Safety Warning" (illuminated at the beginning of the game), a green one labeled "Sequence Loaded," and another green one labeled "Ready/Scanning." Once the tiles are correctly ordered, the "Sequence Loaded" LED turns on. However, the illuminated "MR Safety Warning" LED signals that players must also solve Puzzle 3 before they can proceed.

## Puzzle 3: MR Safety (Scale)

Four ferromagnetic objects need to be found across the room and placed in a box labeled "MR Unsafe", which is placed on top of a scale. Each object is attached to an RF tag (Tabcat, Loc8tor Ltd, London, UK) which can be located with the "ferromagnetic detector" found in the drawer of Puzzle 1. Once the objects are placed in the box on top of the scale, the combined weight is detected and the "scale" device communicates to the electronics of the "sequence composer" to turn off the "MR Safety warning" LED.

## Puzzle 4: Sequence sounds (Speaker)

This is a composite puzzle. Once the “Sequence Loaded” on the “sequence composer” LED is on and the “MR Safety Warning” is off, the “Ready” LED on the “sequence composer” turns green and the pushbutton on the same device can be pressed. The “Ready” LED blinks. While this LED blinks, the “speaker” device is silently activated. This device mimics a standard communication device between the console and the MR scanner room, with a “speaker” button. If the “sequence composer” is not “scanning”, pressing the “speaker” button plays a rhythmic helium pump noise. If the “sequence composer” is “scanning”, pressing the button will let the player hear a series of MR scanner noises corresponding to various conventional acquisition sequences.

## Puzzle 5: Sequence identification and final code (Keypad)

The player is supposed to recognize audio clips of different sequences and match them to their corresponding entry in a "Sequence Handbook" provided in the room. The page number where each sequence is described form a multi-digit code, to be entered into the keypad placed next to the exit door. Entering the correct number triggers a “success” chime, and the timer stops, showing the final time.

# Implementation details

Most of the puzzles were realized through custom electronics, based on the ESP32 microcontroller chip (ESP32-Devkit-C, EspressIF, Shanghai, China) and off-the-shelf components, mounted on custom-manufactured PCB boards and powered through USB power adapters. The firmware code was written with the Arduino platform (Qualcomm, San Diego, USA). All code, schematics, Computer Numerical Control (CNC) machining and 3D printer files, bill-of-materials, and documentation are available at https://github.com/BAMMri/MRPhysics-Escape-Room (version 1.0 [12]) and released under a GNU General Public License.

The communication between the puzzles of the escape room is realized through a serverless hardware network protocol (ESP-NOW [13]) operating on top of the data link layer of a standard Wi-Fi (IEEE 802.11) wireless radiofrequency band. A lightweight, custom communication library was developed to automate mutual detection and online status checking ("heartbeat") of the various components. Through this library, the various components updated each other's statuses directly (for example, the scale could disable the "MR Safety Warning" status of the sequence composer), and sent their own status to the game master's control device. Through this device, the game master was also able to see in real time whether each component was online and react timely to any disruption. A schematic of the communication flow is shown in Figure 2.

## Setup at the ESMRMB Annual Meeting

The escape room was set up in a dedicated area at the 41st ESMRMB Annual Meeting in Marseille (France) in October 2025. Various objects borrowed from the local MR Research Group (such as MR coils, computers, books and folders) were placed on tables and cabinets in order to recreate the atmosphere of an MR control room. A printed poster showing a picture of a 3T MR scanner room was attached to a wall to enhance realism (Figure 3). A webcam was used to stream the videos of the participants inside the room to the game master outside, where the control device was also located. The game master had access to the complete solution and was allowed to give hints to the participants at the participants' request or at the game master's discretion.

The room was open during the regular sessions of the Annual Meeting roughly between 8:30 and 18:30 for three days, so that participants to the meeting could form teams and book a slot, with the frequency of one team every 30 minutes (25 minutes of maximum play time, plus 5 minutes for the game master to reset the room).

No personal information regarding the participants was recorded, only the team name and the solution time in case of success.

# Results

The room worked without technical interruptions throughout the planned duration of operations.

The room was played by 39 teams in total (approximately 160 participants, considering a typical team size of 4). 12 teams solved it (thus resulting in an “escape rate” of 31%), with a median time of 21 minutes and 53 seconds. The fastest solution was in 14 minutes and 10 seconds, whereas the slowest was in 24 minutes and 21 seconds (out of a maximum playing time of 25 minutes).

19 early career researchers were game masters, either alone or sharing their slot with a colleague.

# Discussion

This paper presented an MR-Physics-themed escape room, as it was set up at the 41$^{st}$ Annual Meeting of the ESMRMB held in 2025. Overall, the room was well received by the audience, and the occupancy was full throughout the whole congress.

The playtesting was performed with PhD students and radiographers (MTRA) at the University of Basel, familiar with MR Physics, and the difficulty was adapted according to their feedback. The difficulty of each puzzle was tweaked by adjusting four aspects of each puzzle:

- Objective clarity: how clear it is what is required by the player (example: the player needs to press the “run sequence” button first, and then the “speaker” button).
- Number and availability of components: how easy it is to gather all information/objects required to solve a puzzle (example: the four ferromagnetic objects to put on the scale).

- Prior knowledge: after understanding what the player needs to do, how much outside knowledge is required to solve the puzzle (example: recognizing MR sequences from their noises, which is modulated by the amount of specific information present in the "handbook" given to the player).
- Mental leaps: how many conceptual steps must be taken to go from the solution of one puzzle to the solution of the next, in the same "path" (example: understanding that one must recognize the type of MR acquisition from the sequence sounds, and then finding the corresponding sequence in the "handbook" and understanding that the page number of the handbook corresponds to one digit of the final code). This step can be tweaked by leaving more explicit hints, or inserting smaller puzzles within the same puzzle "path".

The end result was that approximately one third of the participants could solve the room within the allocated time. Although commercial escape rooms can have "escape rates" that vary rather uniformly between less than 10% to more than 90% [14], other escape rooms for similar scientific audiences reported escape rates in line with ours [15, 16]. The target of this room was the audience of a scientific meeting, expected to be trained, on average, at a postgraduate or postdoctoral level, and the teams were usually composed of a mixture of experience levels. Modifying the puzzles by tweaking the above variables could make the room more suitable for a less experienced audience. From observations from the game masters, it was noticed that prior experience both in MR Physics and escape rooms was deemed necessary to complete the game.

Given that gamification has been shown to improve learning outcomes in medical education [6–11] and can promote more positive attitudes toward subjects that learners often perceive as challenging or less engaging [17], a modified version of this escape room could represent a promising supplement to mandatory MR safety training, which is often perceived as repetitive or less motivating, particularly when repeated at regular intervals. The integration of MR safety concepts into collaborative, problem-based activities, as exemplified by the escape room game presented here, has the potential to reinforce fundamental safety principles while concurrently fostering teamwork and communication.

## Limitations

This escape room relies on custom-built, albeit rather simple, electronic components, which, for the most part, cannot be easily repurposed for different scenarios. However, the design concepts and the practical implementation (including the monitoring, and the wireless communication libraries), could be reused and adapted for different embodiments of the game. We are releasing all the implementation details with the highest level of detail possible in the hope that it can be replicated in different contexts, or that it can be of inspiration for similar endeavors in the future.

We have not quantitatively assessed the level of satisfaction of the participants, nor acquired post-game surveys, as the pedagogical value of gamification and escape rooms, especially in the medical field in particular, has been extensively validated in prior research [7, 9, 18, 19]. The purpose of this paper is rather to present the open-source implementation of this specific game.

## Conclusion

This escape room was well received and thoroughly played at the 2025 41st Annual Meeting of the ESMRMB in Marseille (France). We believe that releasing it as open-source in hardware and software can facilitate its deployment in other educational contexts and inspire similar initiatives in the future.

# Acknowledgment

The Authors would like to thank the Local Organizing Committee of the 41st Annual Meeting of the ESMRMB for facilitating the setup of the room during the congress; the Center for Magnetic Resonance in Biology and Medicine (CRMBM) at the Aix-Marseille University (Marseille, France) for providing material and props for the room, specifically Hugo Dary, Aurélien Destruel, Maxime Guye, and Lucas Soustelle; and the Game Masters: Şirin Yağmur Abacı, Lennart Bedarf, Soetkin Beun, Maria Celeste Bonacci, Camilla Calomino, Angelina Catrambone, Ilaria Chimento, Felix Dietz, Martin Freudensprung, Annelie Haek, Leen Hakki, Kato Herregods, Aayush Nepal, Dominique Neuhaus, Judith Schirmer, Valentina Visani, and Simon Weinmüller for their help and dedication.

# Data availability statement

All code, schematics, machining files and documentation of the MRI escape room are released as open source at [12].

# Authors' contributions

Räuber: Study conception and design, Acquisition of Data, Drafting of manuscript

Maggioni: Study conception and design, Acquisition of Data, Critical Revision

Santini: Study conception and design, Drafting of manuscript

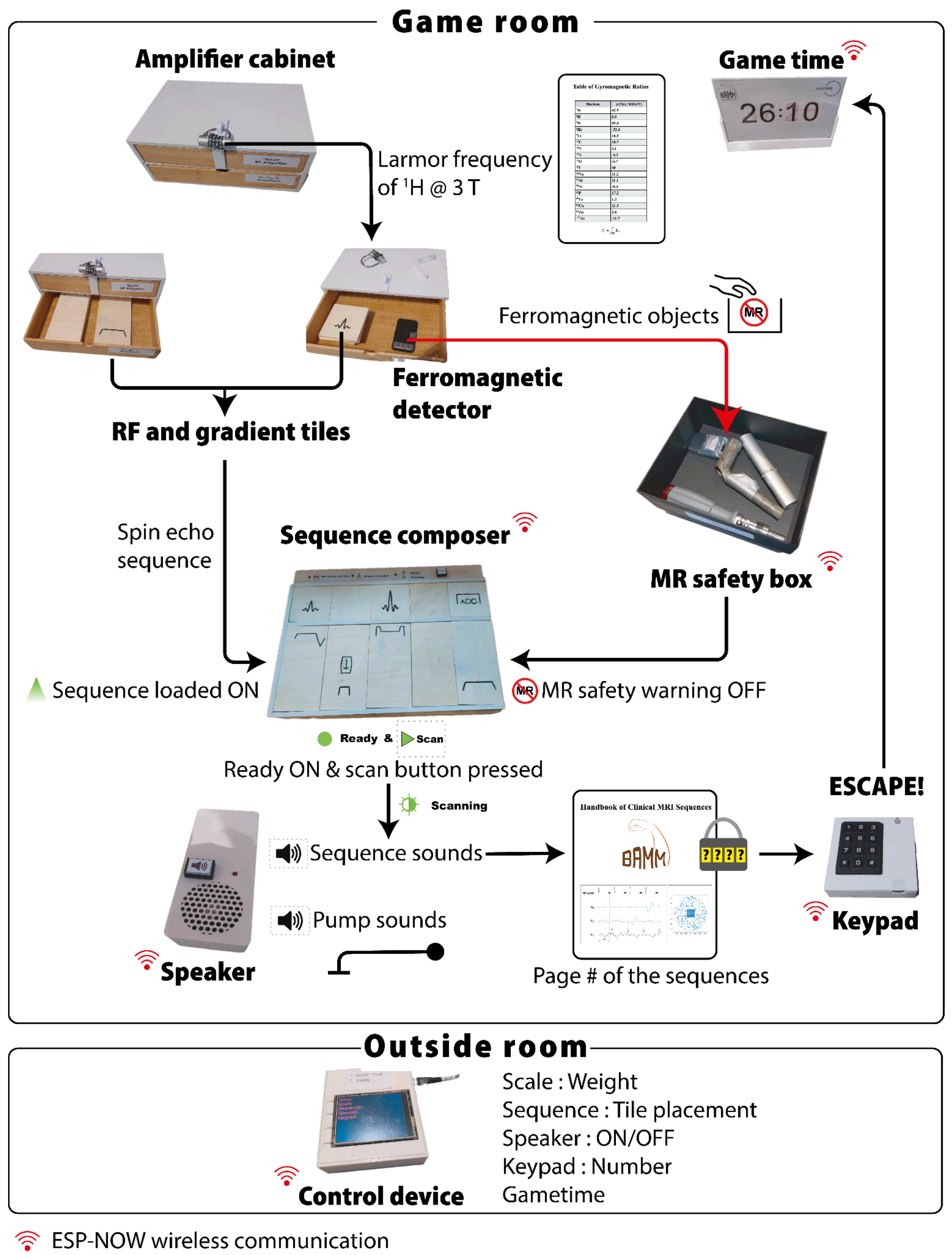

Figure 1: Sketch of the game room layout with puzzles sequence and interactive communication with the gamemaster control device outside of the room.

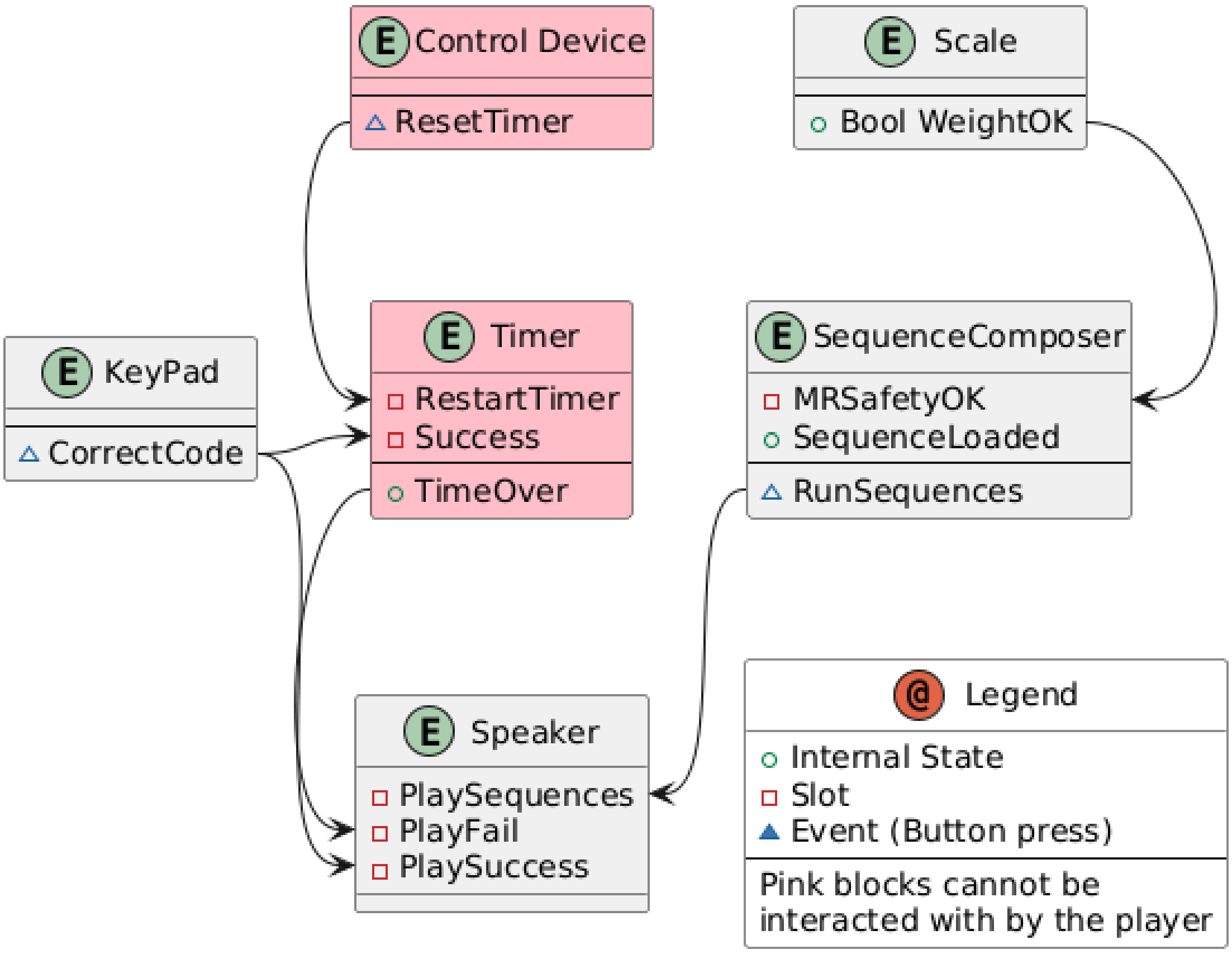

Figure 2: ESP-NOW wireless communication schematics of the MR escape room components. The arrows represent how a status change in one component affects the internal status of another. Every component also sends its status to the control device (not shown in the schematics).

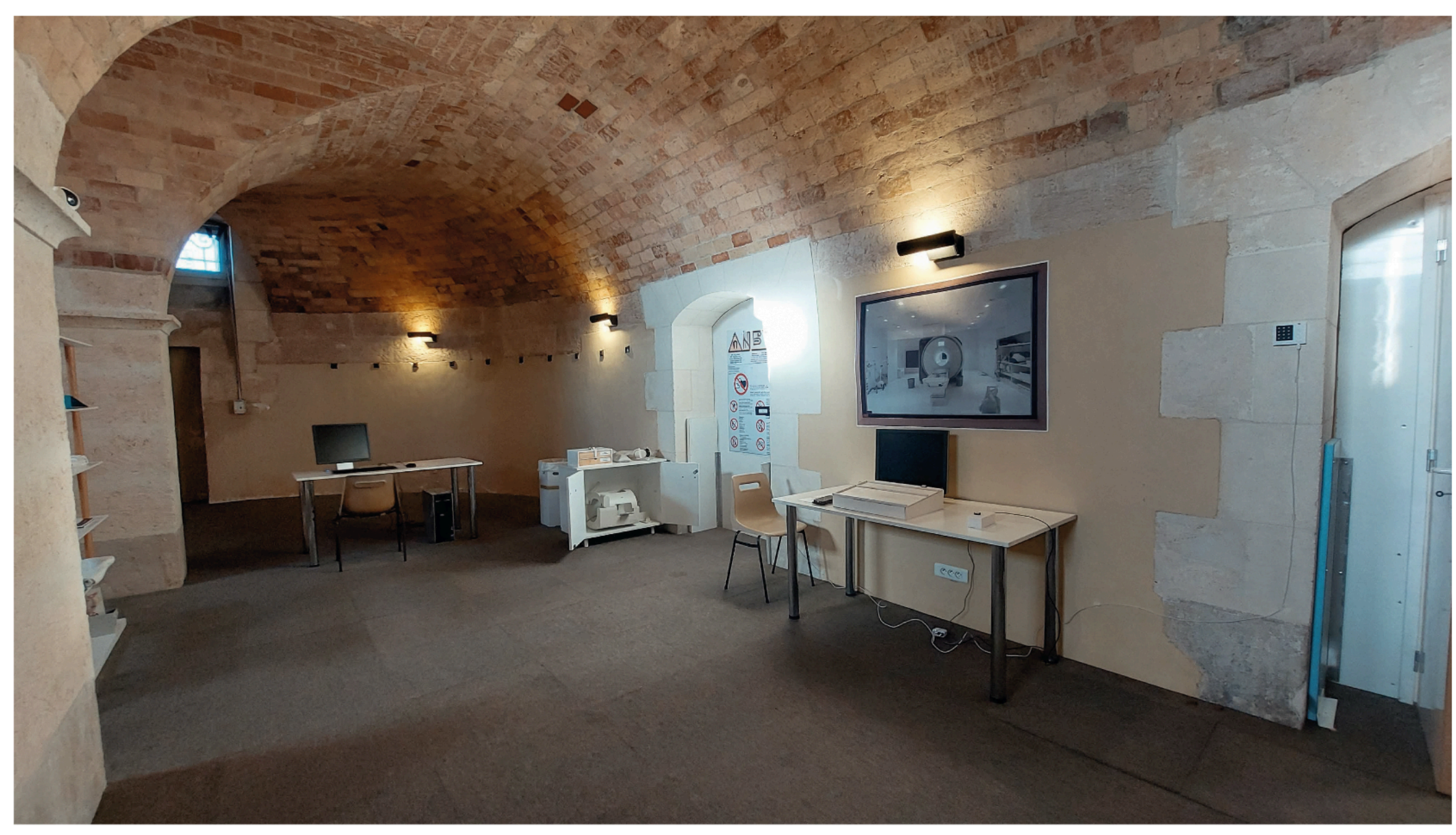

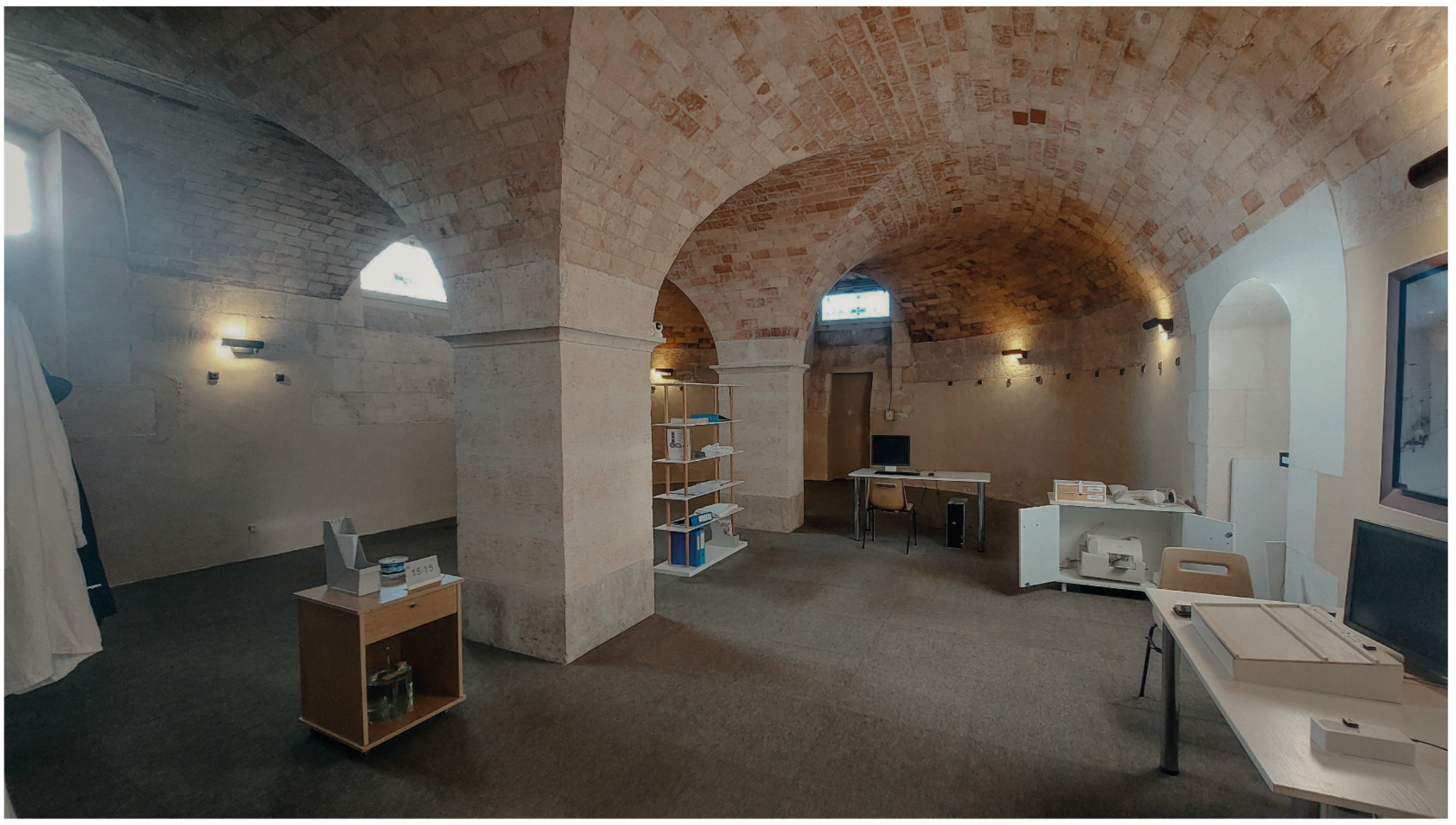

Figure 3: Impressions of the escape room setup at ESMRMB 2025 in Marseille, France.